\documentclass[sigplan,screen]{acmart}
\renewcommand\footnotetextcopyrightpermission[1]{} % removes footnote with conference information in first column
\AtBeginDocument{%
  }

\acmConference[TPSA '25]{}{January 21, 2025}{Denver, CO, USA}

\usepackage{listings}
\usepackage[frozencache,cachedir=.]{minted}
\usepackage{url}
\usepackage{comment}
\usepackage{hyperref}
\usepackage{multirow}
\usepackage{booktabs}
\usepackage{graphicx}
\usepackage{subfig}
\usepackage{float}
\usepackage{amsmath}
\usepackage{algorithm2e}
\RestyleAlgo{ruled}
\SetKwComment{Comment}{/* }{ */}
\definecolor{bg}{rgb}{0.95,0.95,0.95}

\usepackage{cuted}
\usepackage{scalerel}

\renewcommand{\paragraph}[1]{\textbf{#1}}
\usepackage{bm}
\usepackage{caption}
\usepackage{dsfont}

\newcommand{\DFT}{\operatorname{DFT}}

\newcommand{\stride}{\operatorname{L}}

\newcommand{\tensor}{\otimes}

\newcommand{\one}{\operatorname{I}}
\newcommand{\twiddle}{\operatorname{T}}

\definecolor{forestgreen}{HTML}{2E8B57}   % Forest Green
\definecolor{slateblue}{HTML}{6A8EBD}      % Slate Blue
\definecolor{burntorange}{HTML}{C75B1D}    % Burnt Orange
\definecolor{charcoalgray}{HTML}{36454F}

\begin{document}

%%
%% The "title" command has an optional parameter,
%% allowing the author to define a "short title" to be used in page headers.
\title[Towards Semantics Lifting for Scientific Computing]{Towards Semantics Lifting for Scientific Computing: \\ A Case Study on FFT}

%%
%% The "author" command and its associated commands are used to define
%% the authors and their affiliations.
%% Of note is the shared affiliation of the first two authors, and the
%% "authornote" and "authornotemark" commands
%% used to denote shared contribution to the research.
\author{Naifeng Zhang}
\affiliation{%
  \institution{Carnegie Mellon University}
  \city{Pittsburgh}
  % \state{Pennsylvania}
  \country{USA}
}
\email{naifengz@cmu.edu}

\author{Sanil Rao}
\affiliation{%
  \institution{Carnegie Mellon University}
  \city{Pittsburgh}
  % \state{Pennsylvania}
  \country{USA}
}
\email{sanilr@andrew.cmu.edu}

\author{Mike Franusich}
\affiliation{%
  \institution{SpiralGen, Inc.}
  \city{Pittsburgh}
  % \state{Pennsylvania}
  \country{USA}
}
\email{mike.franusich@spiralgen.com}

\author{Franz Franchetti}
\affiliation{%
  \institution{Carnegie Mellon University}
  \city{Pittsburgh}
  % \state{Pennsylvania}
  \country{USA}
}
\email{franzf@andrew.cmu.edu}

%%
%% By default, the full list of authors will be used in the page
%% headers. Often, this list is too long, and will overlap
%% other information printed in the page headers. This command allows
%% the author to define a more concise list
%% of authors' names for this purpose.
% \renewcommand{\shortauthors}{Trovato et al.}

%%
%% The abstract is a short summary of the work to be presented in the
%% article.
\begin{abstract}
The rise of automated code generation tools, such as large language models (LLMs), has introduced new challenges in ensuring the correctness and efficiency of scientific software, particularly in complex kernels, where numerical stability, domain-specific optimizations, and precise floating-point arithmetic are critical. We propose a stepwise semantics lifting approach using an extended SPIRAL framework with symbolic execution and theorem proving to statically derive high-level code semantics from LLM-generated kernels. This method establishes a structured path for verifying the source code's correctness via a step-by-step lifting procedure to high-level specification. We conducted preliminary tests on the feasibility of this approach by successfully lifting GPT-generated fast Fourier transform code to high-level specifications.
\end{abstract}

%%
%% The code below is generated by the tool at http://dl.acm.org/ccs.cfm.
%% Please copy and paste the code instead of the example below.
%%
% \begin{CCSXML}
% <ccs2012>
%  <concept>
%   <concept_id>00000000.0000000.0000000</concept_id>
%   <concept_desc>Do Not Use This Code, Generate the Correct Terms for Your Paper</concept_desc>
%   <concept_significance>500</concept_significance>
%  </concept>
%  <concept>
%   <concept_id>00000000.00000000.00000000</concept_id>
%   <concept_desc>Do Not Use This Code, Generate the Correct Terms for Your Paper</concept_desc>
%   <concept_significance>300</concept_significance>
%  </concept>
%  <concept>
%   <concept_id>00000000.00000000.00000000</concept_id>
%   <concept_desc>Do Not Use This Code, Generate the Correct Terms for Your Paper</concept_desc>
%   <concept_significance>100</concept_significance>
%  </concept>
%  <concept>
%   <concept_id>00000000.00000000.00000000</concept_id>
%   <concept_desc>Do Not Use This Code, Generate the Correct Terms for Your Paper</concept_desc>
%   <concept_significance>100</concept_significance>
%  </concept>
% </ccs2012>
% \end{CCSXML}

% \ccsdesc[500]{Do Not Use This Code~Generate the Correct Terms for Your Paper}
% \ccsdesc[300]{Do Not Use This Code~Generate the Correct Terms for Your Paper}
% \ccsdesc{Do Not Use This Code~Generate the Correct Terms for Your Paper}
% \ccsdesc[100]{Do Not Use This Code~Generate the Correct Terms for Your Paper}

%%
%% Keywords. The author(s) should pick words that accurately describe
%% the work being presented. Separate the keywords with commas.
%% A "teaser" image appears between the author and affiliation
%% information and the body of the document, and typically spans the
%% page.

% \received{20 February 2007}
% \received[revised]{12 March 2009}
% \received[accepted]{5 June 2009}

%%
%% This command processes the author and affiliation and title
%% information and builds the first part of the formatted document.
\maketitle

\section{Introduction}

The growing adoption of neural-based code generation tools, such as large language models (LLMs), presents significant challenges in ensuring the correctness and efficiency of scientific software~\cite{godoy2023evaluation}. Although LLM-generated code may be syntactically valid, it often falls short of meeting the rigorous correctness and performance standards required for complex scientific kernels. Scientific computing demands numerical stability, domain-specific optimizations, and accurate floating-point arithmetic, which are challenging to achieve in code generated without domain expertise. By deriving the semantics of generated kernels statically (at compile time) for cases with unknown (runtime) size parameters, we can identify potential bugs, inefficiencies, and performance bottlenecks before deploying the code. This statically derived information can also be fed back into neural-based code generation tools to iteratively improve the generated code.
However, scientific computing poses unique challenges for static analysis tools. Accurate handling of floating-point arithmetic requires managing rounding errors and numerical precision, while pointers, recursion, and transcendental functions like sine and cosine further complicate static analysis. To address these issues, this work proposes \emph{stepwise semantics lifting} as an early-stage experimental solution within constrained boundaries. We develop a novel extension to the SPIRAL system~\cite{puschel2005spiral,franchetti2018spiral}, which is equipped with symbolic execution and theorem-proving capabilities. Through an LLVM-to-SPIRAL parser, we import LLM-generated scientific kernels into SPIRAL and derive their semantics using SPIRAL's formal framework and engine.

\paragraph{Contributions.} To summarize, this paper makes the following contributions:
\vspace{-1mm}
\begin{enumerate}
    \item An experimental approach, stepwise semantics lifting, for statically extracting high-level semantics from scientific kernels.
    \item An end-to-end demonstration of the proposed approach by lifting GPT-generated fast Fourier transform code to its high-level specification.
\end{enumerate}
\vspace{-3mm}

\section{Background}

In this section, we provide background on SPIRAL, a formal code generation system, and the target scientific kernel: the fast Fourier transform (FFT).

\paragraph{The SPIRAL system.}
\label{sec:bg-spiral}
The SPIRAL system~\cite{puschel2005spiral} originated as an automatic \sloppy{performance-tuning} system for signal processing algorithms, particularly focusing on FFT algorithms. This focus stemmed from the availability of a formal framework (the Kronecker product formalism~\cite{johnson1990methodology,van1992computational}), which enables capturing and manipulating FFT algorithms in high-level mathematical representations. Over time, this representation was generalized to encompass a broader range of algorithms~\cite{franchetti2009operator}, including both sparse and dense mathematical computations. SPIRAL has thus evolved into a comprehensive code generation system, capable of taking high-level specifications and producing optimized implementations for target platforms.

\paragraph{SPIRAL dialects.} The SPIRAL system consists of three main components used in a stepwise code generation process: i) Signal Processing Language (SPL)~\cite{xiong2001spl}, ii) $\Sigma$-SPL~\cite{franchetti2005formal}, and iii) internal code (icode)~\cite{franchetti2018spiral}. SPL, the top-level domain-specific language (DSL), describes the mathematical semantics of kernels and the functional data flow of the target algorithm. The lower-level DSL, $\Sigma$-SPL, captures loop abstractions, while icode serves as an abstract code representation adaptable to different code syntaxes. As shown in Figure~\ref{fig:overview}, each layer of abstraction is connected by a rewrite system that applies recursive descent followed by confluent term rewriting. For further details, we refer readers to the respective citations.

\paragraph{Formal guarantees of SPIRAL.} SPIRAL is built on top of the GAP computer algebra system~\cite{GAP4}, enabling localized correctness checks during rule application between abstraction layers. This allows verification that the left-hand and right-hand sides of each rewrite rule are equivalent at every step. Previous work, namely HELIX~\cite{zaliva2018helix}, has demonstrated that SPIRAL's algebraic guarantees can be extended to provide theorem prover-level assurances.

\paragraph{The FFT algorithm.}
The discrete Fourier transform (DFT) is a fundamental tool in science and engineering, playing a key role in areas such as signal processing, spectral analysis, communications, machine learning, and finance. 
The FFT is a class of efficient algorithms for computing the DFT. While a direct computation of the DFT for an \(n\)-element vector requires \(O(n^2)\) operations, the FFT reduces this complexity to \(O(n \log n)\). 

\paragraph{FFT algorithms in SPL.} In SPIRAL, linear transforms are represented as matrix-vector multiplications. For example, the DFT definition is viewed as a matrix-vector product:
\begin{equation}
\label{eq:dft-mv}
    y = \DFT_n x, \quad \DFT_n = [\omega^{kl}_n]_{0 \leq k,l < n},
\end{equation}
where $\omega_{n}=e^{-2\pi i / n}$ and $i = \sqrt{-1}$. 
There are several FFT algorithms for computing the DFT. 
Using SPL, we can define one of the most widely adopted FFT algorithms, the recursive Cookley-Tukey FFT algorithm, as 
\begin{equation}
\label{eq:fft-ct}
    \DFT_n = (\DFT_m \tensor \one_{k}) \twiddle^n_{k} (\one_m \tensor \DFT_{k}) \stride^n_m, \ n = mk,
\end{equation}
where $\one$ is the identity matrix, $\twiddle$ is the twiddle matrix and $\stride$ is the stride permutation matrix. $\stride^{mk}_m$ permutes the elements of the input vector as $im + j \mapsto jk + i, 0\leq i < k, 0 \leq j < m$~\cite{franchetti2011fast}.

\section{Stepwise Semantics Lifting}

\begin{figure}[t]
\centering
    \includegraphics[width=0.47\textwidth,trim={0 3mm 0 0},clip]{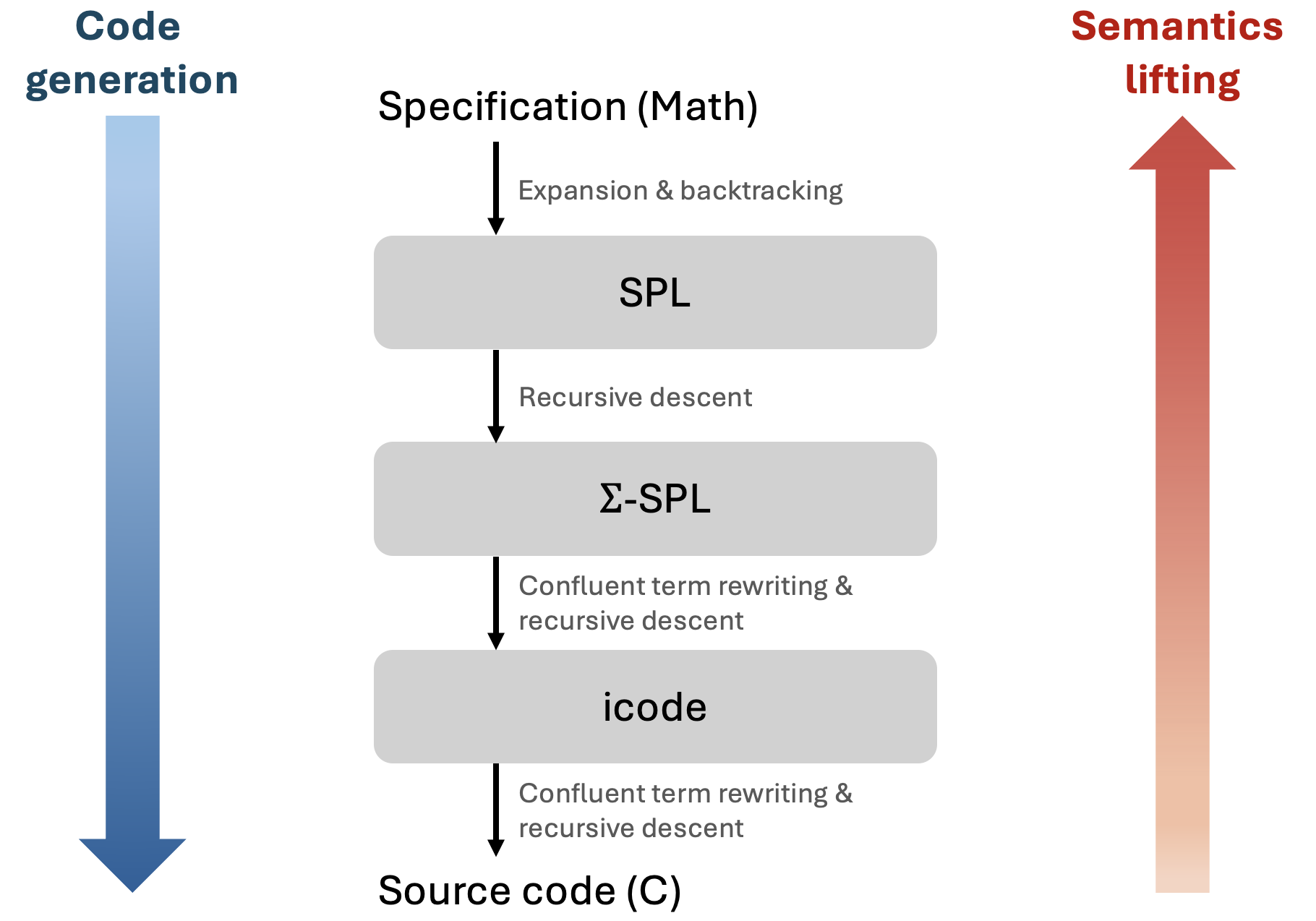}
    \caption{Overview of proposed semantics lifting procedure via SPIRAL. 
    We propose to reverse the well-established code generation (i.e., lowering) process~\cite{puschel2005spiral,franchetti2018spiral,xiong2001spl,franchetti2005formal,low2017high} to stepwisely lift the semantics of the source code.
    }
    \label{fig:overview}
\end{figure}

Our proposed approach explores the feasibility of lifting scientific kernels by reversing SPIRAL’s rule-based code generation (i.e., lowering) process to transform source code into the highest achievable abstraction level, which is illustrated in Figure~\ref{fig:overview}. 
Rather than employing linear rewriting as in the lowering process, lifting is structured as a search problem, where each step identifies the sequence of rule applications necessary to increase the abstraction level.

We demonstrate our approach through an end-to-end lifting of FFT code generated by GPT-4~\cite{achiam2023gpt}, as presented in Listing~\ref{lst:fft-c}. 
We have developed a custom parser that extracts the code’s abstract syntax tree (AST) and converts it to SPIRAL’s intermediate representation, icode. We omit the discussion of this parser due to space constraints.

\vspace{-1mm}
\begin{listing}[!htb]
\begin{minted}[frame=none, obeytabs=true, tabsize=4, linenos, numbersep=-6pt, escapeinside=||, fontsize=\tiny]{c}
    #include <math.h> 
    #include <stdlib.h>
    #define M_PI 3.14159265358979323846
    void fft_recursive(double* data, int n) {
        if (n <= 1) return;
        // Allocate temporary storage for half-size FFTs.
        double* even = (double*)malloc(n * sizeof(double));
        double* odd = (double*)malloc(n * sizeof(double));
        for (int i = 0; i < n / 2; ++i) {
            even[2 * i] = data[4 * i];
            even[2 * i + 1] = data[4 * i + 1];
            odd[2 * i] = data[4 * i + 2];
            odd[2 * i + 1] = data[4 * i + 3];
        }
        // Recursively compute FFTs.
        fft_recursive(even, n / 2); 
        fft_recursive(odd, n / 2);
        for (int i = 0; i < n / 2; ++i) {
            double theta = -2.0 * M_PI * i / n;
            double wr = cos(theta);
            double wi = sin(theta);
            // Twiddle factor multiplication.
            double real = odd[2 * i] * wr - odd[2 * i + 1] * wi;
            double imag = odd[2 * i] * wi + odd[2 * i + 1] * wr;
            data[2 * i] = even[2 * i] + real;
            data[2 * i + 1] = even[2 * i + 1] + imag;
            data[2 * (i + n / 2)] = even[2 * i] - real;
            data[2 * (i + n / 2) + 1] = even[2 * i + 1] - imag;
        }
        // Cleanup.
        free(even); 
        free(odd);
    }
\end{minted}
\caption{GPT-generated recursive FFT in C, which implements Equation~\ref{eq:fft-ct}.}
\label{lst:fft-c}
\end{listing}
\vspace{-1mm}

\paragraph{Internal code to $\Sigma$-SPL.}
The provided example contains two loop bodies. We will now demonstrate how to lift lines 9-14 to $\Sigma$-SPL. 
To capture the access patterns in the code block, we utilize the gather and scatter formalism within SPIRAL~\cite{low2017high}. In general, any read and write access can be formalized as an outer product of an $n \times 1$ and a $1 \times n$ standard basis vector: 
$$
e^{n \times 1}_{s(j)} \cdot e^{1 \times n}_{g(j)},
$$
where $e^{n \times 1}_{k}$ (resp. $e^{1 \times n}_{k}$) is an $n \times 1$ (resp. $1 \times n$) vector with a one at the $k^\text{th}$ position and zeros elsewhere.
For the next step, we first need to conduct a range analysis on \verb|data|, which is a pointer passed into the function without explicitly specified ranges. 
Now, line 10 turns into 
$$
e^{n\times 1}_{2j} \cdot e^{1\times 2n}_{4j}.
$$
We apply a similar analysis to lines 11-13 to obtain $B_j$, and then we can write the entire loop (lines 9-14) as
\begin{equation}
\label{eq:sspl}
    \sum_{j=0}^{n/2-1} B_j, \quad B_j = \sum_{i=0}^{3} e^{n\times 1}_{2j+(i \operatorname{mod} 2)} \cdot e^{1\times 2n}_{4j+i}.
\end{equation}
Equation~\ref{eq:sspl} can be directly captured by the iterative sum operator in $\Sigma$-SPL~\cite{franchetti2005formal}.

\paragraph{$\Sigma$-SPL to SPL.}
To lift $\Sigma$-SPL to SPL for the given example, we recognize the pattern of interleaved real and imaginary formats in this step, as $B_j$ gathers from the same source and writes to two outputs twice, with an index offset of $1$. The interleaved format of complex numbers is represented by the $\overline{(\cdot)}$ operator in SPIRAL~\cite{franchetti2002simd} (which corresponds to \verb|RC()| in Listing~\ref{lst:fft-splfunc}). Therefore, we can constrain the search problem to permutations of a complex vector. By definition, a stride-2 permutation reads with a stride of two and writes continuously, which aligns with the behavior observed in our target code block (lines 10-13). Consequently, the output from this stage is 
\begin{equation}
    \overline{\stride^n_2},
\end{equation}
which, in SPL, represents a stride-2 permutation for a vector of $n$ complex numbers. 

\paragraph{Induction on SPL objects.} After lifting all loop bodies to SPL expressions, we need to combine all lifted SPL components to form a complete SPL expression. In our example, we demonstrate how to lift lines 10-13 to $\overline{\stride^n_2}$; similar principles apply to lifting lines 18-29, which results in $\overline{(\DFT_2 \tensor \one_{n/2}) \twiddle^n_{n/2}}$. Therefore, we lift the code in Listing~\ref{lst:fft-c} to the representation in Listing~\ref{lst:fft-splfunc}.

\vspace{-1mm}
\begin{listing}[!htb]
\begin{minted}[frame=none, obeytabs=true, tabsize=4, linenos, numbersep=-6pt, escapeinside=||, fontsize=\tiny]{c}
    void fft_recursive(double* data, int n) {
        |\textcolor{slateblue}{splfunc(concat\_array(even, odd), data, i, RC(L(n, 2)));}|
        |\textcolor{burntorange}{fft\_recursive(even, n / 2);}|
        |\textcolor{burntorange}{fft\_recursive(odd, n / 2);}|
        |\textcolor{forestgreen}{splfunc(data, concat\_array(even, odd), i,}|
        |\textcolor{forestgreen}{    RC(Tensor(F(2), I(n/2)) * Diag(n, n/2)));}|
    }
\end{minted}
\caption{GPT-generated recursive FFT partially represented by SPL.}
\label{lst:fft-splfunc}
\end{listing}
\vspace{-1mm}

Since the FFT is defined recursively in this example, in SPL, we can write Listing~\ref{lst:fft-splfunc} as 
\begin{equation}
\label{eq:spl-induction}
    \textcolor{burntorange}{M_n} = \textcolor{forestgreen}{\overline{(\DFT_2 \tensor \one_{n/2}) \twiddle^n_{n/2}}} \textcolor{burntorange}{(\one_2 \tensor \;M_{n/2})} \textcolor{slateblue}{\overline{\stride^n_2}},
\end{equation}
where $M_n$ is an $n \times n$ matrix. 

Now we can symbolically execute the entire program by setting $n = 2$. As FFT is a linear transform, we can derive the transformation matrix by combining the coefficients of each element in the result vector. By equivalence matching this matrix, we can verify that the generated C code implements a $\overline{\DFT_2}$ when $n = 2$. Given the base case, we can then induct on Equation~\ref{eq:spl-induction} and derive that $M_n = \overline{\DFT_n}$. 
Therefore, we can consolidate all components into a single SPL expression:
\begin{equation}
\label{eq:spl-derived}
    \textcolor{forestgreen}{\overline{(\DFT_2 \tensor \one_{n/2}) \twiddle^n_{n/2}}} \textcolor{burntorange}{\overline{(\one_2 \tensor \DFT_{n/2})}} \textcolor{slateblue}{\overline{\stride^n_2}}.
\end{equation}
Note that the SPL expression, by definition, is evaluated from right to left. Thus, Equation~\ref{eq:spl-derived} aligns precisely with the sequence of operations in Listing~\ref{lst:fft-c} and~\ref{lst:fft-splfunc}.

\paragraph{SPL to mathematical specification.}
This marks the final step of the lifting process. In this phase, we match the derived SPL expression with existing entries in SPIRAL's knowledge base through pattern matching. SPIRAL contains a wide range of linear transform algorithms, particularly within the FFT family~\cite{franchetti2018spiral}. 
In our example, the derived SPL (Equation~\ref{eq:spl-derived}) corresponds to Equation~\ref{eq:fft-ct} when $m = 2$, $k = n/2$, and inputs are complex numbers. 
Hence, we can conclude that, with computer algebra system-level guarantees, the source code successfully implements the Cooley-Tukey FFT algorithm recursively. We write the final output as $\DFT_n: \mathbb{C}^n \mapsto \mathbb{C}^n.$

\paragraph{Towards lifting multilinear operations.}
Our proposed approach can be extended to other scientific kernels, such as axpy~\cite{blackford2002updated}, a multilinear operation that takes in two vectors of the same length, scales one vector by a constant $\alpha$ and adds it pointwise to the other vector. 
We can represent axpy for two vectors of length $n$ in $\Sigma$-SPL as follows:
\begin{equation}
\label{eq:axpy-sspl}
    e_j^{n\times 1} \cdot [1, \alpha] \cdot \one_2 \tensor \; e^{n \times 1}_j.
\end{equation}
We can then identify the scaling and reduction patterns within the axpy operation and lift Equation~\ref{eq:axpy-sspl} to SPL as 
\begin{equation}
    [\one_n | \; \alpha \one_n].
\end{equation}

\section{Related Work}

The field of verified lifting~\cite{kamil2016verified,qiu2024tenspiler} utilizes SMT solvers to search for and verify program summaries (in the form of loop invariants and post-conditions) that correspond to the source code. However, existing efforts in this domain, which focus on stencil and tensor computations, fall short for scientific kernels due to limitations with floating-point numbers, pointers, and recursive functions~\cite{qiu2024tenspiler}.
Our approach shares the spirit of Stephen Wolfram's integration of Wolfram|Alpha with GPT~\cite{wolfram2023chatgpt}, but leverages a more robust domain theory~\cite{franchetti2009operator} suited to the complexity of floating-point numerical software, a domain typically hard to standard formal methods~\cite{barrett1989formal,harrison2006floating}. Chelini et al.~\cite{chelini2020multilevel} similarly attempt to reverse the compilation process for scientific applications but restrict their approach to lifting matrix multiplications using the \verb|Affine| dialect to \verb|Linalg| dialect in MLIR.

\section{Conclusion}

In this work, we propose stepwise semantics lifting for scientific kernels and demonstrate a preliminary end-to-end lifting from LLM-generated C code to high-level semantics that recognizes the source code as a recursive FFT implementation. The problem of deriving the formal semantics of numerical code may indeed be unsolvable in general. However, for the class of algorithms that the SPIRAL system addresses—a well-defined subset of floating-point computational science and engineering algorithms—the problem becomes tractable. 
We aim to evaluate the proposed approach on a wider range of algorithms, focusing on numerical routines with data-independent control flows. Additionally, enhancing LLM code generation for scientific kernels by integrating SPIRAL-lifted information with human-guided prompts~\cite{godoy2024large} could be a promising future direction.

%%
%% The acknowledgments section is defined using the "acks" environment
%% (and NOT an unnumbered section). This ensures the proper
%% identification of the section in the article metadata, and the
%% consistent spelling of the heading.
\newpage
\begin{acks}
This material is based upon work supported by the National Science Foundation under Grant No. 1127353 and the U.S. Department of Energy, Office of Science, Office of Advanced Scientific Computing Research under Award Number DE-FOA-0002460. Any opinions, findings, and conclusions or recommendations expressed in this material are those of the authors and do not necessarily reflect the views of the National Science Foundation and the U.S. Department of Energy. Franz Franchetti was partially supported as the Kavčić-Moura Professor of Electrical and Computer Engineering.
\end{acks}

%%
%% The next two lines define the bibliography style to be used, and
%% the bibliography file.
\bibliographystyle{acmref}
\bibliography{ref}

%%
%% If your work has an appendix, this is the place to put it.
% \appendix

\end{document}